\documentclass[useAMS,usenatbib]{mn2e}
\usepackage{graphicx}
\usepackage{color}
\newcommand{\ion}[2]{#1\,{\sc #2}}
\newcommand{\etal}{et al.}

\title[On The Relationship Between Magnetic Cancellation And UV Burst Formation]{On The Relationship Between Magnetic Cancellation And UV Burst Formation}
\author[C. J. Nelson \etal]{C. J. Nelson$^{1,2,3}$\thanks{E-mail:
c.j.nelson@sheffield.ac.uk}, J. G. Doyle$^{2}$ \& R. Erd\'elyi$^{1,4}$\\
$^{1}$School of Mathematics and Statistics, Hicks Building, University of Sheffield, Hounsfield Road, Sheffield , S3 7RH, UK\\
$^{2}$Armagh Observatory, College Hill, Armagh, BT61 9DG, Northern Ireland, UK\\
$^{3}$Astrophysics Research Centre, School of Mathematics and Physics, Queen's University, Belfast, BT7 1NN, Northern Ireland, UK\\
$^{4}$Debrecen Heliophysical Observatory (DHO), Hungarian Academy of Science, 4010 Debrecen, P.O. Box 30, Hungary}

\begin{document}

\pagerange{\pageref{firstpage}--\pageref{lastpage}} \pubyear{2016}

\maketitle

\label{firstpage}

\begin{abstract}
Burst-like events with signatures in the UV are often observed co-spatial to strong line-of-sight photospheric magnetic fields. Several authors, for example, have noted the spatial relationship between Ellerman bombs (EBs) and Moving Magnetic Features (MMFs), regions of flux which disconnect from a sunspot or pore before propagating away in the moat flow and often displaying evidence of cancellation. In this article, data collected by the Solar Dynamics Observatory's {\it Helioseismic and Magnetic Imager} and {\it Atmospheric Imaging Assembly} are analysed in an attempt to understand the potential links between such cancellation and UV burst formation. Two MMFs from AR $11579$, three bi-poles from AR $11765$, and six bi-poles (four of which were co-spatial to IRIS bursts) in AR $11850$ were identified for analysis. All of these cancellation features were found to have lifetimes of the order hours and cancellation rates of the order $10^{14}$-$10^{15}$ Mx s$^{-1}$. H$\alpha$ line wing data from the Dunn Solar Telescope's {\it Interferometric BIdimensional Spectrometer} were also available for AR $11579$ facilitating a discussion of links between MMFs and EBs. Using an algebraic model of photospheric magnetic reconnection, the measured cancellation rates are then used to ascertain estimates of certain quantities (such as up-flow speeds, jet extents, and potential energy releases) which compared reasonably to the properties of EBs reported within the literature. Our results suggest that cancellation rates of the order measured here are capable of supplying enough energy to drive certain UV bursts (including EBs), however, they are not a guaranteeing condition for burst formation.
\end{abstract}


\section{Introduction}

Small-scale burst-like features with signatures in the UV spectrum have been widely researched in the literature. Ellerman Bombs (EBs; \citealt{Ellerman17}), for example, are small-scale brightening events, originally observed in the wings of the H$\alpha$ line profile, which, until recently (see \citealt{vanderVoort16}), were thought to form uniquely within Active Regions (ARs). Recent researches have also, however, associated EBs with $1600$ \AA\ and $1700$ \AA\ continuum intensity increases as well a sub-set of the recently discovered IRIS bursts (see, {\it e.g.}: \citealt{Peter14}; \citealt{Vissers15}). EBs often manifest as elongated structures with a major axis length of around $1$\arcsec\ and exhibit dynamic fine sub-structuring (as was discussed by \citealt{Watanabe11}, \citealt{Nelson15}). The lifetimes of EBs are of the order minutes (see, for example, \citealt{Zachariadis87}, \citealt{Georgoulis02}, \citealt{Watanabe11}), although, short-lived intense increases in brightness and area (presented by \citealt{Qiu00}, \citealt{Watanabe11}, \citealt{Vissers15}) have been suggested to be indicative of a high-energy driver, widely hypothesised to be photospheric magnetic reconnection (recently researched by, {\it e.g.}, \citealt{Georgoulis02}, \citealt{Vissers13}, \citealt{Nelson13b}).

\begin{figure*}
{\includegraphics[scale=1.47]{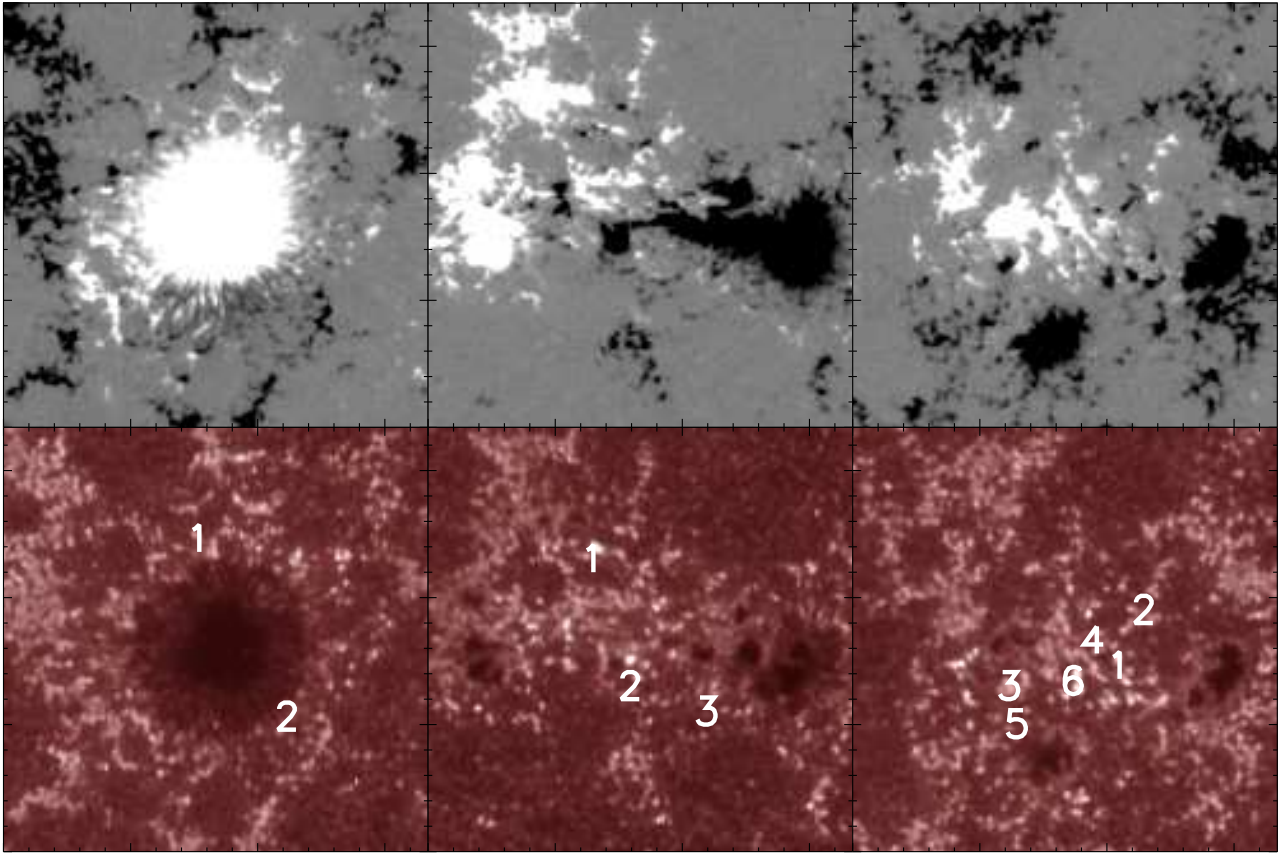}}
\caption{The $100$\arcsec$\times100$\arcsec\ FOV of each of the three ARs analysed in this study sampled at their respective initial time-step by the SDO/HMI instrument (top row) and the SDO/AIA $1700$ \AA\ filter (bottom row). Positive polarity is indicated by white pixels and negative polarity is indicated by black pixels. Each of the magnetograms and each of the $1700$ \AA\ images are artificially saturated at $\pm300$ G and $6000$ counts, respectively, for ease of comparison. The ordering of the ARs from left to right is: AR $11579$; AR $11765$; AR $11850$. Approximate spatial locations of the UV bursts discussed in this article are indicated by the white numbers in the SDO/AIA $1700$ \AA\ panels.}
\label{overview_anal}
\end{figure*}

Photospheric magnetic reconnection could prove to be an important process in driving numerous features from the lower solar atmosphere into the outer reaches of the Sun, including certain spicules (\citealt{dePontieu07}), filaments (\citealt{Litvinenko99b}, \citealt{Litvinenko07}), surges (\citealt{Roy73}), and the recently discovered IRIS bursts (\citealt{Peter14}). \citet{Georgoulis02} presented three cartoon magnetic field topologies, one of which agrees well with the expected magnetic topology of Moving Magnetic Features (MMFs; \citealt{Sheeley69}, \citealt{Harvey73}, \citealt{Lim12}), which could facilitate the formation of UV bursts (assuming these features were a product of magnetic reconnection). Indeed, correlations between EBs and MMFs have been found (see, for example, \citealt{Nindos98}) meaning the relationship between these features could be worthy of continued study. One potential avenue of research, followed in this article, is the measurement and subsequenst analysis of the often cited cancellation rates of MMFs (recently researched by \citealt{Li13}) and other bi-polar regions co-spatial with a range of UV bursts.

The development of analytical tools which have attempted to model photospheric Sweet-Parker reconnection (\citealt{Parker57}; \citealt{Sweet58}) by \citet{Litvinenko99a} and \citet{Litvinenko07} have now made the measurement of such cancellation rates co-spatial to a variety of UV bursts desirable. In this article, the observed properties of cancelling bi-poles are recorded before being inputted into the model developed by \citet{Litvinenko99a} in order to assert whether realistic outputs, comparable with observations of UV bursts (for example, the upward flow velocity of EBs measured to be $9$ km s$^{-1}$ observed by \citealt{Nelson15}), are returned. Such comparisons are important for testing both the models and the magnetic reconnection hypothesis as the driver of certain UV bursts.

The method developed by \citet{Litvinenko99a} has so far provided interesting results with regards to filament and EUV jet formation (see, for example, \citealt{Litvinenko99b}, \citealt{Chae03}) and could, therefore, shed light on the sporadic relationship between EBs and surges reported in the literature (see, for example, \citealt{Rust68}, \citealt{Yang13}, \citealt{Reid15}). Specifically, it is of interest to compare any measured cancellation rates co-spatial to UV bursts to those observed co-spatial to surges (\citealt{Roy73}) and EBs (\citealt{Reid16}). If magnetic reconnection were to be the reason for the observed cancellation, it is possible that the minimum cancellation rate for EBs to form is lower than the minimum required for surge formation, potentially explaining the plethora of observations depicting EBs with no links to surges. 

\begin{table*}
\begin{center}
\begin{tabular}{|c|c|c|c|c|}
\hline
Start time (UT) & End time (UT) & Routine & Frames & Cadence \\
\hline
$14$:$34$:$19$ & $14$:$49$:$20$ & B & $133$ & $6.8$ seconds \\
$14$:$51$:$04$ & $15$:$05$:$55$ & A & $162$ & $5.4$ seconds \\
$15$:$07$:$35$ & $15$:$22$:$32$ & C & $377$ & $2.4$ seconds \\
$15$:$24$:$12$ & $15$:$39$:$13$ & B & $133$ & $6.8$ seconds \\
$15$:$41$:$00$ & $15$:$59$:$59$ & A & $207$ & $5.4$ seconds \\
$16$:$01$:$53$ & $16$:$15$:$17$ & C & $338$ & $2.4$ seconds \\
$16$:$17$:$49$ & $16$:$32$:$50$ & B & $133$ & $6.8$ seconds \\
\hline
\end{tabular}
\caption{Summary of ground-based observations available for AR $11579$ with routines `A', `B', and `C' defined as the H$\alpha$ line scan, $30$-frame speckle, and $10$-frame speckle sequences (described in more detail in the text), respectively. It should be noted that the initial implementation of sequence `A' corresponds to the data analysed in \citet{Nelson13b} (with a slight difference in the timing due to the discounting of time between the sequence initialisation and data acquisition and the inclusion of four frames at the end of the sequence which were removed from the previous article due to poor seeing).}
\label{times}
\end{center}
\end{table*}

It should be noted, however, that this study does not aim to directly address the links between UV bursts and surges. Any process which could potentially link such events is still open to discussion and would likely require the combination of observations with realistic numerical simulations beyond the scope of the present article. A number of researches have, though, hypothesised that shock waves, excited by a reconnection event such as an EB in the lower solar atmosphere, could act to raise the chromospheric-coronal separating layer, hence leading to the ejection of mass as a surge (see, for example, \citealt{Suematsu82}, \citealt{Shibata82}, \citealt{Takasao13}).

The work contained within this article is structured as follows: In Section $2$, the relevant observations are presented. The identification of relevant cancellation features, their basic properties, and their links to UV bursts are described in Section $3$, before the findings are incorporated into the models of \citet{Litvinenko99a} and \citet{Litvinenko07} in Section $4$. Finally, a discussion about the relevance of these results is conducted in Section $5$.

\section{Observations}

\subsection{AR $11579$}

The first field-of-view (FOV) discussed in this article was sampled within NOAA AR $11579$ on the $30$th September $2012$ between $14$:$00$ UT and $17$:$00$ UT. Ground-based data were collected using the {\it Interferometric BIdimensional Spectrometer} (IBIS; \citealt{Cavallini04}) instrument at the Dunn Solar Telescope (DST) with an $80$\arcsec\ circular diameter centred on the leading sunspot of the AR, intially situated at co-ordinates of $x_\mathrm{c}\approx28$\arcsec, $y_\mathrm{c}\approx-275$\arcsec\ (with respect to the disc centre; $\mu$=$0.96$). The spatial resolution of these data is approximately $0.2$\arcsec. Co-spatial line-of-sight photospheric magnetic fields were inferred using Solar Dynamics Observatory's {\it Helioseismic and Magnetic Imager} (SDO/HMI; \citealt{Scherrer12}) data, which were downloaded, reduced, and cropped to a $100$\arcsec$\times100$\arcsec\ box centred on the DST/IBIS FOV. Co-alignment was achieved by matching bright regions in the H$\alpha$ line wings (hypothesised to be a good proxy of the vertical magnetic field by, for example, \citealt{Rutten13}) with small-scale magnetic field elements in the SDO/HMI images. Finally, data collected by the $1700$ \AA, $1600$ \AA, and $304$ \AA\ filters on the Solar Dynamics Observatory's {\it Atmospheric Imaging Assembly} (SDO/AIA; \citealt{Lemen12}) were downloaded and aligned to provide information about the local UV and EUV signals. The initial magnetic configuration of this FOV and the co-spatial $1700$ \AA\ structuring are plotted in the first column of Fig.~\ref{overview_anal} for reference. Overall, this region remained relatively stable through time, however, a steady stream of MMFs were observed to flow away from the sunspot, many of which exhibited cancellation.

Three distinct observational routines (which shall be denoted as `A', `B', and `C' for ease) were employed by the DST/IBIS instrument over a two-hour period between $14$:$34$:$19$ UT and $16$:$32$:$50$ UT. Routine `A' comprised a $17$-point H$\alpha$ line scan and a $9$-point \ion{Fe}{I} $6302.5$ \AA\ line scan, neither of which sampled with even spacing through the profiles. The minimum and maximum wavelengths of the observed H$\alpha$ line profile were $-0.99$ \AA\ and $+1.01$ \AA\ from $6562.8$ \AA, respectively. The \ion{Fe}{I} $6302.5$ \AA\ data are not analysed in this article and so will not be discussed further. In total, $26$ images were taken per sequence repetition with a cadence of approximately $5.4$ seconds. Routine `B' imaged the wings of the H$\alpha$ line profile at only $-0.74$ \AA\ and $+0.76$ \AA, acquiring $30$ frames at each line position with a total of $60$ frames per repetition. Each set of $30$ frames was then reduced using the speckle method (see \citealt{Woger08}) returning data with a total cadence of around $6.8$ seconds. Finally, routine `C' is similar to routine `B' except that each line position was only sampled $10$ times, giving a total of $20$ frames per repetition at a cadence of $2.4$ seconds.  A summary of the two-hour period of observations can be found in Table~\ref{times}.

\subsection{AR $11765$}

The second region of interest to this study is AR $11765$, which has previously been shown to contain around $20$ EBs during the hour-long observations analysed by \citet{Hong14}. A $100$\arcsec$\times100$\arcsec\ FOV (initially centred on co-ordinates $x_\mathrm{c}\approx-142$\arcsec\ and $y_\mathrm{c}\approx143$\arcsec; $\mu=0.98$) was isolated and tracked for a two hour period between $16$:$00$ UT and $18$:$00$ UT on the $6$th June $2013$ (encompassing the temporal coverage researched by \citealt{Hong14}) when the region was still in its emerging phase. This region contained numerous sunspots/pores of both positive and negative polarity in addition to a high number of network elements and dynamic burst-like events, which appeared to show significant evolution of their structuring through this two hour period. As with AR $11765$, line-of-sight magnetic fields were inferred using the SDO/HMI instrument and the co-spatial UV/EUV emission was inferred using SDO/AIA $1700$ \AA, $1600$ \AA, and $304$ \AA\ data. The majority of burst-like events within this FOV occurred co-spatial to apparent horizontal structuring (potentially similar to fibrils commonly observed within the H$\alpha$ line core) observed within the $304$ \AA\ filter which appear to cover the centre of the AR. The initial line-of-sight magnetic field and $1700$ \AA\ filter response within this FOV are plotted in the middle column of Fig.~\ref{overview_anal}.

\subsection{AR $11850$}

The third region used as a test for this article is AR $11850$. This AR was of interest for this research due to the existence of IRIS UV bursts co-spatial to measurable cancellation, discussed by \citet{Peter14}. A $100$\arcsec$\times100$\arcsec\ FOV (original co-ordinates: $x_\mathrm{c}\approx-271$, $y_\mathrm{c}\approx59$; $\mu=0.96$) was selected and tracked for a two-hour period between $11$:$00$ UT and $13$:$00$ UT on the $24$th September $2013$. Several sunspots/pores are evident within the FOV, in addition to a variety of flux emergence and cancellation events (some of which are co-spatial to the events analysed by \citealt{Peter14}) typical of an emerging AR. The initial line-of-sight magnetic field configuration and $1700$ \AA\ SDO/AIA filter response co-spatial to this FOV are plotted in the right hand column of Fig.~\ref{overview_anal}.

\begin{figure}
\begin{center}
{\includegraphics[scale=0.33,trim={0 7.5cm 0 1.8cm}]{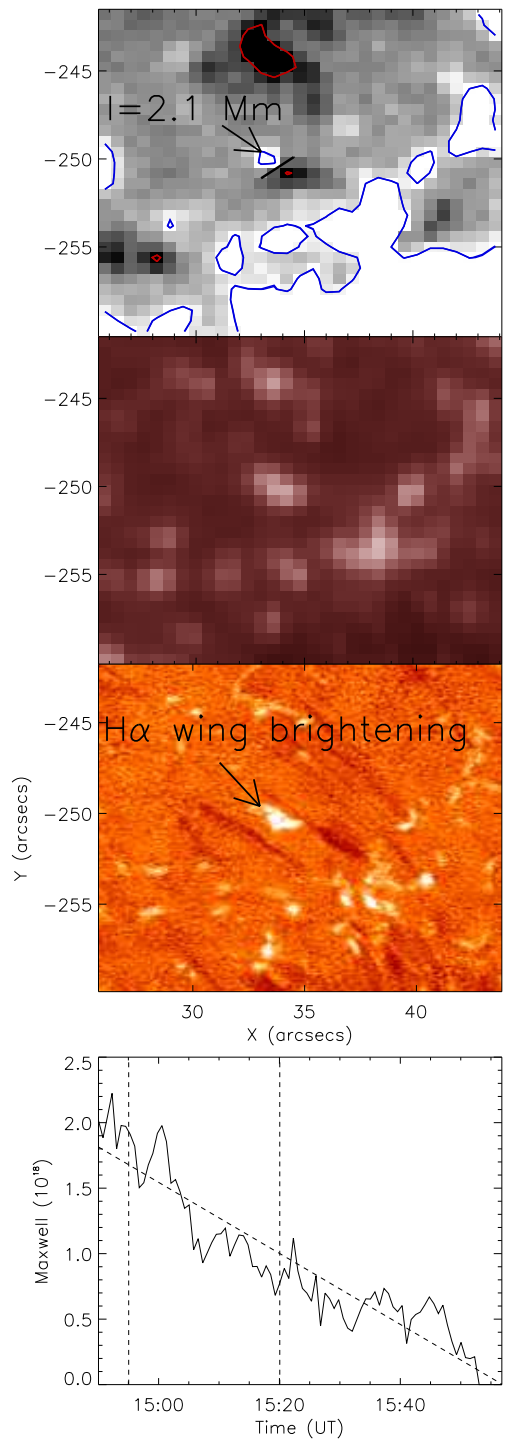}}
\caption{The FOV surrounding MMF $1$ at approximately $15$:$15$:$33$ UT. The top row plots the SDO/HMI sampled line-of-sight magnetic field artificially saturated at $\pm 100$ G. The blue (positive polarity) and red (negative polarity) contours outline the pixels above this threshold and the black line indicates the cross-sectional diameter. The co-spatial SDO/AIA $1700$ \AA\ filter image saturated at $8000$ counts (second row) depicts limited intensity increases, however, the H$\alpha$ blue wing saturated at $150$ \% of the background intensity ($-0.74$ \AA; third row) displays higher intensity increases. The bottom row plots the magnetic flux through time for MMF $1$, with the first and second dashed lines indicating the onset and disappearance of the intensity increase in the H$\alpha$ line wings, respectively.}
\label{caneve_anal}
\end{center}
\end{figure}

\section{Data Analysis}

\subsection{Tracking and Feature Identification}

\begin{figure*}
{\includegraphics[scale=0.32]{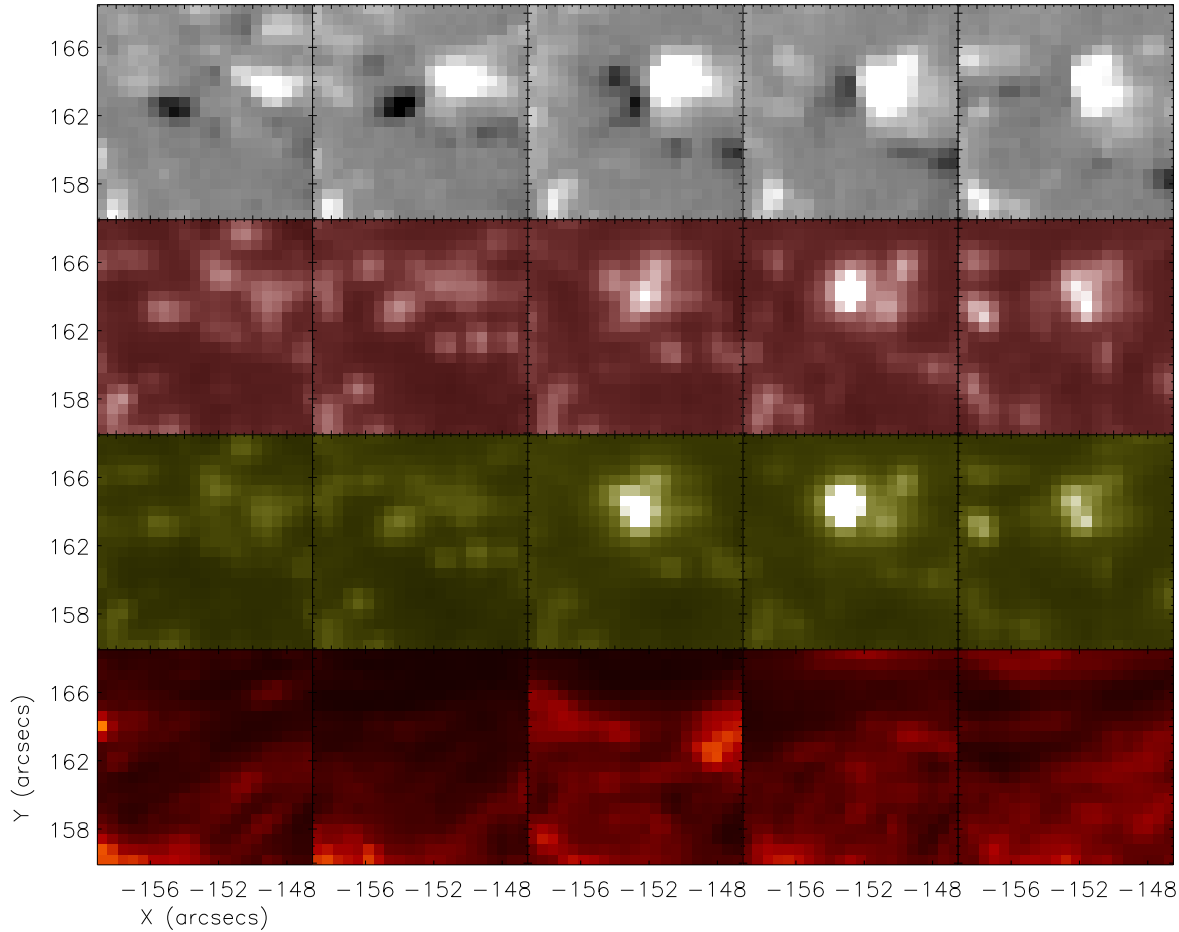}}
\caption{The evolution of Can. Event $1$ from AR $11765$. The negative polarity field can be seen to cancel against the region of positive polarity in the SDO/HMI magnetograms (top row) which occurs co-spatial to increases in intensity within the SDO/AIA $1700$ \AA\ and $1600$ \AA\ filters (second and third rows, respectively). The SDO/AIA $304$ \AA\ filter shows no evidence of a co-spatial or co-temporal response to this localised cancellation. The images in each row are plotted with an artificial saturation level of $\pm300$ G, $8000$ counts, $2500$ counts, and $2000$ counts (from top to bottom, respectively). The left-hand column is plotted at the closest time-step to $16$:$23$:$09$ UT. Each subsequent column is separated by approximately $22$ minutes.}
\label{evolution_anal}
\end{figure*}

As the main focus of this article is on the relationship between cancellation features and UV bursts, it was first necessary to identify a method to track potential bi-poles which were sufficiently isolated through time such that accurate measurements of their properties ({\it e.g.}, magnetic field strength) could be made. The tracking of magnetic features within these datasets was completed using {\it Yet Another Feature Tracking Algorithm} (YAFTA; \citealt{Welsch03}), which required minimum thresholding of both magnetic field strength and area as inputs. In this study, these thresholds were set at $40$ G and $2$ pixels ($0.5$\arcsec$\times1$\arcsec), respectively, through a systematic testing of a variety of parameters with the aim of securing a successful and consistent identification of small-scale flux elements. The magnetic field strength, area, and cross-sectional diameter (taken to be length of the `current sheet' in Section 4) of all regions of interest could then be recorded for each relevant frame and saved in individual files for further analysis. 

After the tracking method had been finalised, numerous examples of cancellation were identified within each FOV (including features co-spatial to the potential EBs discussed in \citealt{Nelson13b} [AR $11579$] and IRIS bursts discussed by \citealt{Peter14} [AR $11850$]). These cancellation features could be split into three individual categories, namely: ($1$) Isolated regions of cancellation spatially separated from any other measured flux; ($2$) Cancellation which merged with other small regions of flux during the lifetime of the bi-pole; ($3$) Cancellation of one small, uni-polar flux region against a larger body of opposite polarity network. Intuitively, features within the second category must be disregarded for this study due to a masking of the cancellation by the merging flux. Overall, eleven representative cancellation features were identified within these datasets for analysis, including two small-scale MMFs from AR $11579$ (not co-spatial to those analysed by \citealt{Nelson13b}), three larger cancelling bi-poles from AR $11765$, and six bi-poles (four of which were co-spatial to the IRIS bursts discussed by \citealt{Peter14}) in AR $11850$. In order to be consistent throughout our analysis, a single representative polarity was selected for each event and measured through time in order to infer the cancellation rate of the region.

Following the detection of cancellation features, the co-spatial and co-temporal SDO/AIA $1600$ \AA\ and $1700$ \AA\ imaging data were studied to confirm whether a UV burst was present (threshold values of $2500$ counts and $8000$ counts, respectively, which were deemed suitable across all datasets). Nomenclature defining specific UV bursts is an important and a non-trivial matter, with a variety of features being associated with comparable increases in emission. EBs for example, especially for relatively disk-centre observations such as those presented here (in order to gain a high-accuracy magnetic field dataset), can be difficult to distinguish from magnetic concentrations in the H$\alpha$ line wings (as has recently been discussed by \citealt{Rutten13}, \citealt{Vissers13}, \citealt{Vissers15}). Here, therefore, the focus will be on physical quantities of any UV bursts (such as area, lifetime, intensity increase) which can be compared across these datasets as well as back to typical values of EBs in the literature. It should be noted that no direct attribution of these UV features to previously discussed events (such as EBs) will be attempted and that we shall stick to the more general term of UV bursts. Finally, the SDO/AIA $304$ \AA\ data were studied to determine whether any co-spatial EUV signal was present.

\subsection{AR $11579$}

The first feature analysed in this article (hereafter referred to as MMF $1$) propagated away from the North-West portion of the penumbra of the sunspot between $14$:$45$ UT and $16$:$00$ UT. Prior to the measured cancellation, a large region of positive polarity flux fragmented, before one of the segments interacted with a region of emerging negative polarity flux. Both regions of polarity then cancelled sufficiently through time such that they were no longer observable in SDO/HMI magnetogram data. The morphological properties of MMF $2$ (identifiable South-East of the sunspot between $15$:$50$ UT and $16$:$40$ UT) were similar to those demonstrated by MMF $1$. Initially, a region of positive polarity flux was observed to fragment from a larger body close to the sunspot before interacting with an emerging negative polarity flux segment. 

These identified MMFs were typical examples of such features (as originally discussed by \citealt{Sheeley69} and within the large statistical sample analysed by \citealt{Li13}). Both MMF $1$ and MMF $2$ had cross-sectional diameters of around $2$-$3$ Mm, original flux measurements of the order $10^{18}$ Mx, and lifetimes of around one hour meaning any properties derived from these features should be suitable to be extrapolated to a larger body of events. In Fig.~\ref{caneve_anal}, the FOV co-spatial to MMF $1$ is plotted for the SDO/HMI line-of-sight magnetic field (top row), the SDO/AIA $1700$ \AA\ filter (second row; artifical threshold of $6000$ counts), and the H$\alpha$ blue wing (third row; saturated at $150$ \% of the background intensity) depicting the identified bi-pole and apparent H$\alpha$ intensity increase. In the bottom panel, the magnetic flux is also plotted highlighting an obvious decrease through time. The black line in the top frame indicates the estimated cross-sectional diameter and the dashed lines in the bottom frame indicate the onset and disappearance of any observed H$\alpha$ intensity increases (not necessarily above a $150$ \% value). Measuring the fluxes of these MMFs through time, it was possible to find cancellation rates, $R$, of $4.1\times10^{14}$ Mx s$^{-1}$ and $6.4\times{10}^{14}$ Mx s$^{-1}$ for MMFs $1$ and $2$, respectively. Approximate widths of the bi-poles, $l$, are found to be $3.1$ Mm and $2.1$ Mm, again respectively. In the following sections, these values shall be put in context, however, initially a brief discussion of the imaging data co-spatial to these events will be conducted.

\begin{figure*}
{\includegraphics[scale=0.3]{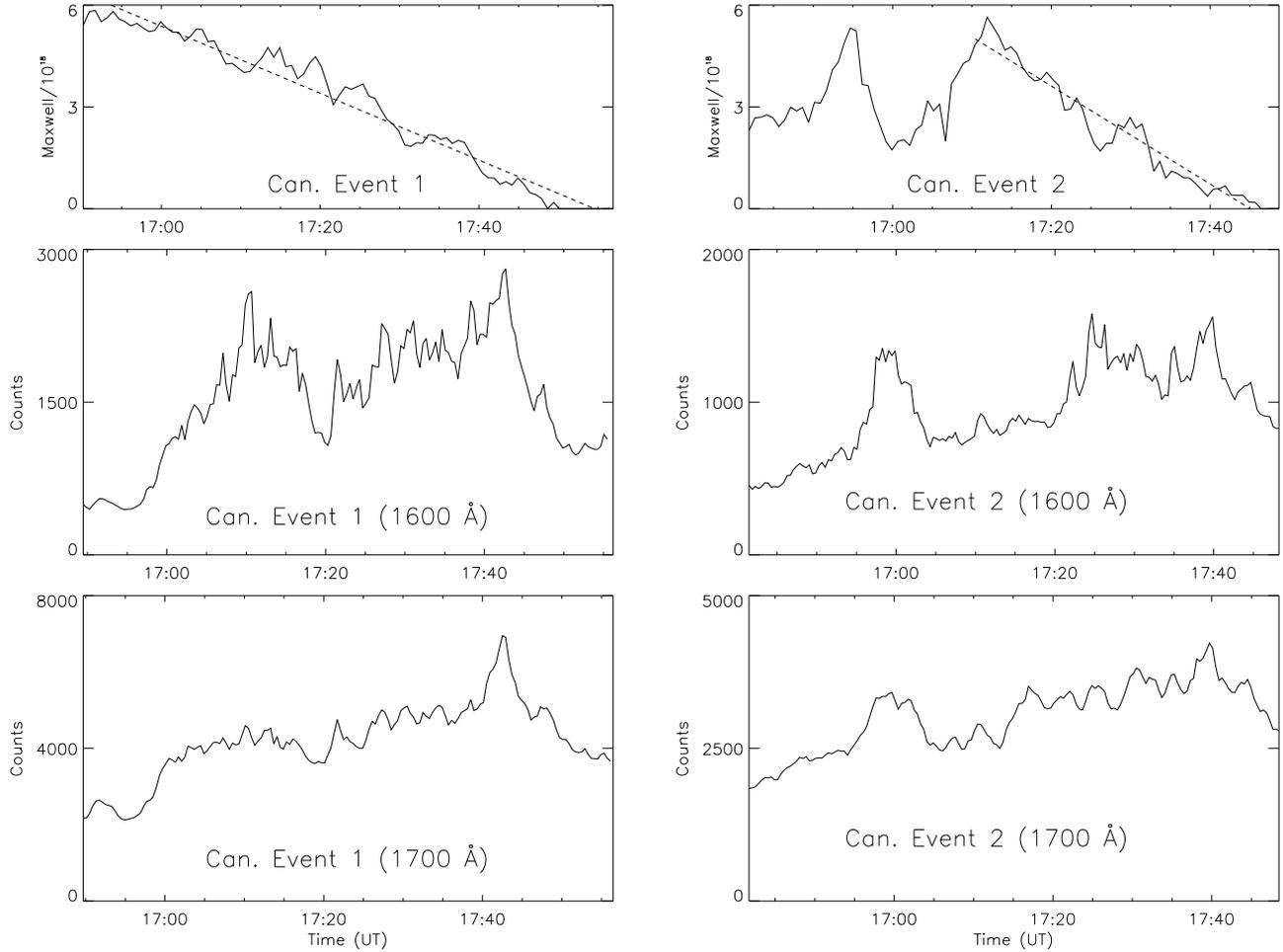}}
\caption{(Top row) Measurements of the total negative polarity magnetic field contained within two bi-poles identified in AR $11765$ (indicated in the individual frames). The dashed lines overlay the linear fit used to estimate the cancellation rate for each event. (Middle row) SDO/AIA $1600$ \AA\ and (Bottom row) SDO/AIA $1700$ \AA\ lightcurves averaged over a $7\times7$ pixel box surrounding the UV burst for the duration of the cancellation. These average intensities are often lower than the thresholds defined previously.}
\label{can_11765}
\end{figure*}

The H$\alpha$ emission features co-spatial to both MMFs $1$ and $2$ were, in many ways, similar to EBs, with diameters of slightly over $1$\arcsec, lifetimes of approximately $10$ minutes, and occasional increases in intensity over $150$ \% of the background intensity (comparable to events discussed by, for example, \citealt{Zachariadis87}, \citealt{Georgoulis02}, \citealt{Nelson15}). Initially, both features manifested as relatively small, weak brightenings in the H$\alpha$ line wings, before displaying dynamical morphological evolutions. It is interesting to note that for both features, the H$\alpha$ wings only exhibited emission for a small section of time during which cancellation was observed. This increased intensity corresponded well to the period of greatest cancellation although no firm conclusions can be drawn on the relevance or generality of this due to our small sample size of MMFs.

Whether these increases in the H$\alpha$ wing emission are EBs or the more common magnetic concentrations (as were discussed by \citealt{Rutten13}) is difficult to ascertain due to the changing intensity thresholds from observing sequence to sequence. As this FOV includes only limited flux emergence and a light flow of MMFs, it is unfortunately also not possible to quantify the percentage of MMFs which correspond to H$\alpha$ intensity increases (such as was conducted by \citealt{Nindos98}). We do note that each of the further obvious bi-polar MMFs (which fell into category $2$; {\it i.e.}, interacted with apparently emerging flux during their lifetimes) observed within these SDO/HMI data corresponds to at least some limited brightening activity in the H$\alpha$ line wings, but whether this was due to the increased H$\alpha$ wing emission associated with magnetic concentrations or due to the occurrence of EBs is unknown. A longer continuous observing sequence should be analysed in more detail in future research in order to further understand this.

Recent work by \citet{Vissers13} and \citet{Vissers15} has indicated that identifying burst-like behaviour in the SDO/AIA UV channels is an excellent method for confirming detections of EBs. The $1600$ \AA\ and $1700$ \AA\ emissions co-spatial to these cancellation events were, therefore, examined in an attempt to draw further inferences about the nature of the H$\alpha$ line wing intensity increases. The SDO/AIA UV emission co-spatial and co-temporal to these MMFs were relatively limited, with some short-lived increases in intensity but no major burst-like emission observed (over the thresholds discussed previously). It is, therefore, possible that any increases in emission within the H$\alpha$ line co-spatial to these cancellation events is related to the more benign magnetic concentrations (see, for example, \citealt{Rutten13}) rather than explosive EBs. In addition to this, no signature of burst-like events was observed in the SDO/AIA $304$ \AA\ channel, although this would also be expected for EBs (see, for example, \citealt{Vissers13}, \citealt{Nelson15}).

\subsection{AR $11765$}

The FOV studied within AR $11765$ appeared to be a more dynamic environment then AR $11579$ with a higher number of flux emergence and cancellation events occurring within the analysed two-hour time period. The first cancellation event discussed here (referred to as Can. Event $1$) was observed at initial (from the original frame) co-ordinates of $x_\mathrm{c}\approx-152$\arcsec, $y_\mathrm{c}\approx162$\arcsec\ between $16$:$50$ UT and $17$:$57$ UT. Two regions of flux, one positive and one negative, began to approach one another at around $16$:$20$ UT before the negative polarity field fully cancelled away. The evolution of this feature is plotted in Fig.~\ref{evolution_anal} for reference. The second feature (Can. Event $2$) suitable for analysis within these data occurred at co-ordinates of approximately $x_\mathrm{c}\approx-144$\arcsec, $y_\mathrm{c}\approx129$\arcsec\ between $17$:$12$:$39$ UT and $17$:$49$:$24$ UT. A large ($5$ Mm in diameter) region of positive polarity flux was observed to move towards a stable region of negative polarity field before rapidly cancelling away. The final bi-pole of interest in AR $11765$ (Can. Event $3$) occurred in an apparently more benign region of this FOV at initial co-ordinates of $x_\mathrm{c}\approx-125$\arcsec, $y_\mathrm{c}\approx129$\arcsec\ between $16$:$05$ UT and $17$:$45$ UT. A segment of negative polarity field disconnected from a larger body before cancelling with a small region of positive polarity flux.

All three cancellation features analysed within AR $11765$ were larger in area than those discussed in the previous subsection, with approximate peak cross-sectional diameters close to $5$ Mm and initial flux estimates over $5\times10^{18}$ Mx. Can. Event $1$ had a similar lifetime to typical MMFs (see, for example, \citealt{Li13}), however, Can. Event $2$ was shorter lived, occurring for close to $40$ minutes. Can. Event $3$, in turn was slightly longer lived. Interestingly, all three events displayed cancellation rates an order of magnitude larger than the features analysed in AR $11579$ with values of $1.6\times10^{15}$ Mx s$^{-1}$, $2.4\times10^{15}$ Mx s$^{-1}$, and $1.6\times10^{15}$ Mx s$^{-1}$ measured for Can. Event $1$, Can. Event $2$, and Can. Event $3$, respectively. The estimated cross-sectional diameters of these events during the cancellation were $2.6$ Mm, $2.5$ Mm, and $2.3$ Mm. The evolution of the magnetic field contained within the negative polarity poles of Can. Events $1$ and $2$ is plotted in the top row of Fig.~\ref{can_11765} and strongly depicts this cancellation. 

Analysis of the SDO/AIA $1600$ \AA\ and $1700$ \AA\ filters highlighted strong increases in intensity co-spatial and co-temporal to both Can. Event $1$ (as is displayed in Fig.~\ref{evolution_anal}) and Can. Event $2$. Indeed, both of these interacting bi-poles occurred co-spatial to one of the two most intense features observed within this FOV during this time-period by the $1700$ \AA\ filter. The UV burst co-spatial to Can. Event $1$ occurred between $17$:$00$ UT and $17$:$50$ UT with a cross-sectional diameter of the order $4$ Mm, and exhibited several impulsive brightenings followed by a short fading. The UV burst co-spatial to Can. Event $2$, however, had a peak width of around $5$ Mm and was much more dynamic with several apparent kernels forming over numerous impulsive phases. Lightcurves for $7\times7$ pixel boxes encompassing these events are plotted in the middle and bottom rows of Fig.~\ref{can_11765}. Interestingly, when the lightcurves for Can. Event $2$ are compared to the evolution of the magnetic field plotted in the top row of Fig.~\ref{can_11765}, each sustained intensity increase appears to be linked to a period of strong cancellation. One obvious example of this is the UV burst at approximately $17$:$00$ UT in the lightcurve for Can. Event $2$. Overall, these lightcurves appear to match the signatures detected by \citet{Vissers15} co-spatial to EBs and Flaring Arch Filaments (FAFs). 

Interestingly, no UV or EUV burst was detected co-spatial to Can. Event $3$ despite the comparable cancellation rates. A number of explanations are available to account for the lack of observed UV burst including, but not limited to, the low viewing angle masking the formation of any event against the background magnetic concentrations (as has been discussed by \citealt{Rutten13}, \citealt{Vissers13}), that this event forms as a sub-class of cancellation which does not display strong evidence of a co-spatial UV burst, and that any burst behaviour occurred below the temporal resolution of the SDO/AIA UV filters. A larger sample of cancellation features, co-temporally sampled in the H$\alpha$ line wings, would be required to further investigate any such possibilities.

\begin{figure*}
{\includegraphics[scale=0.3]{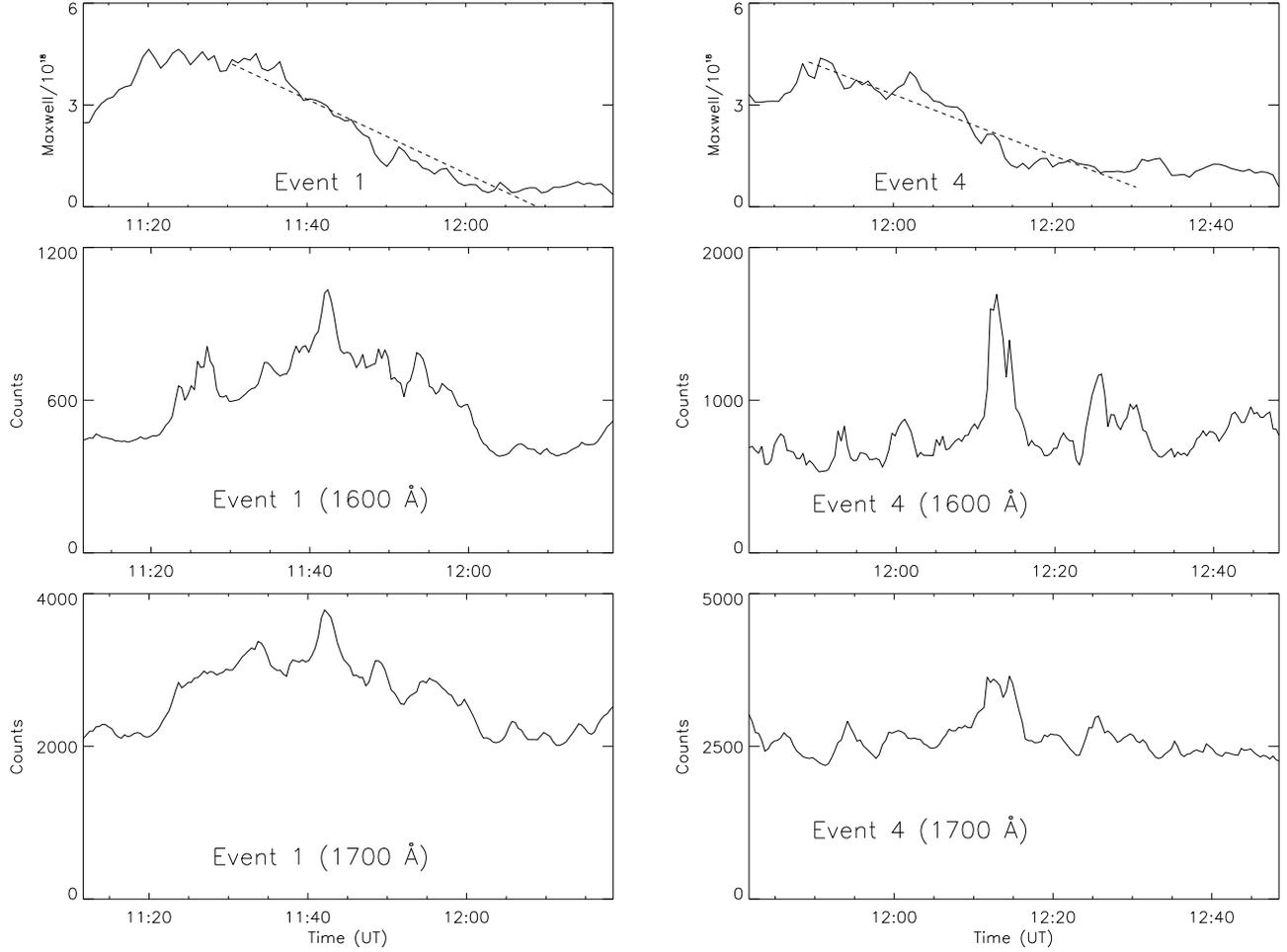}}
\caption{Same as Fig.~\ref{can_11765} but for two representative events from AR $11850$.}
\label{can_11850}
\end{figure*}

\subsection{AR $11850$}

During the two-hour period analysed in this article, the FOV studied within AR $11850$ was also very dynamic with numerous flux emergence and cancellation events being observed. \citet{Peter14} discussed four IRIS bursts which occurred between $11$:$44$ UT and $12$:$04$ UT co-spatial to bi-polar magnetic fields. Over the course of the two hours, all four of these regions clearly show some level of cancellation meaning they are suitable for analysis here. We shall, therefore, adopt the same numbering convention for these events as \citet{Peter14}.

For these UV bursts, cancellation rates can be derived using the method previously implemented on both AR $11579$ and AR $11765$. Values for the cancellation rate, $R$, were found to be $1.8\times10^{15}$ Mx s$^{-1}$, $1.5\times10^{15}$ Mx s$^{-1}$, $3.7\times10^{15}$ Mx s$^{-1}$, and $1.5\times10^{15}$ Mx s$^{-1}$ for events $1$ to $4$, respectively. These measurements were made for the positive polarity flux for events $2$ and $3$ and the negative polarity flux for events $1$ and $4$. Overall, these values correspond well to the cancellation rates measured for Can. Events $1$, $2$, and $3$ within AR $11765$. Cross-sectional diameters were estimated to be $2.2$ Mm, $2.6$ Mm, $3.5$ Mm, and $2.2$ Mm, once again for events $1$ to $4$, respectively. The negative polarity flux contained within bi-polar Events $1$ and $4$ is plotted through time in the top row of Fig.~\ref{can_11850}.

Of these four cancelling bi-poles, three (events $1$, $3$, and $4$) formed co-spatial to strong UV bursts in both the $1600$ \AA\ and $1700$ \AA\ filters similar to the first two events discussed within AR $11765$. Of these events, the burst corresponding to event $3$ appeared to be the weakest with only short-lived measureable intensity enhancements, observed early on during the analysed time-period. The UV bursts co-spatial to events $1$ and $4$, however, were repetitive, with numerous impulsive brightening phases observed during the course of these data. Average lightcurves (again calculated for a $7\times7$ pixel box) surrounding these UV bursts are plotted in the middle and bottom rows of Fig.~\ref{can_11850}. These lightcurves illustrate the strong impulsive behaviour observed, specifically within the $1600$ \AA\ filters.

The UV continuum co-spatial to Event $2$ displayed no such behaviour, with no intensity increases above the current thresholds. In terms of cancellation rates and $304$ \AA\ response, no difference between this event and the other features analysed in for this FOV is discernible. It is possible, as with the FAFs discussed by \citet{Vissers15}, that the energy deposition relating to this event could occur higher in the atmosphere but still below the chromospheric canopy, limiting both the photospheric ($1700$ \AA) and chromospheric ($304$ \AA) response. This lack of UV burst co-spatial to the IRIS burst is interesting and deserving of future research, potentially conducted using co-spatial IRIS spectra and magnetograms collected using the {\it CRisp Imaging SpectroPolarimeter} (CRISP; \citealt{Scharmer06}, \citealt{Scharmer08}).

In addition to the features studied in detail by \citet{Peter14}, six further UV bursts were identified during this time period, each of which was co-spatial to a bi-pole. Four of these features occurred within the confines of the region of apparent flux emergence at the centre of the FOV and were, therefore, not suitable for analysis in this article. Interestingly, one of these events is clearly visible in the SDO/AIA $1600$ \AA\ and $1700$ \AA\ channels during the time period analysed by \citet{Peter14}, however, no IRIS burst was identified indicating either no increased \ion{Si}{IV} $1394$ \AA\ signal was produced at this site or that any heightened emission was missed by the raster. Cancellation rates, $R$, were measured for the two remaining (short-lived) events (denoted as Events 5 \& 6) at $5\times10^{14}$ Mx s$^{-1}$ and $1.9\times10^{15}$ Mx s$^{-1}$. The widths of the bi-poles were estimated to be $1.3$ Mm and $2.2$ Mm, respectively. For reference, Event $5$ (Event $6$) occurred at initial co-ordinates of $x_\mathrm{c}\approx-285$\arcsec\ ($\approx-267$\arcsec), $y_\mathrm{c}\approx39$\arcsec\ ($\approx51$\arcsec).

\section{Model Predictions}

\subsection{Application of the Model}

This section focuses on the the initial model of \citet{Litvinenko99a} and the slightly expanded version set out by \citet{Litvinenko07}. For ease of the reader these models shall be briefly described here. This analysis assumes that a vertical current sheet exists in the photosphere between two opposite polarity regions. The length, thickness, and height of the current sheet are defined as $l$, $2a$, and $2b$ (where $b\approx\Lambda$, the atmospheric scale height). The plasma densities (temperatures) at the entrance to the current sheet and within the current sheet are denoted as $n_\mathrm{i}$ ($T_\mathrm{i}$) and $n$ ($T$). In addition to this, the magnetic field strength and in-flow speed at the entrance to the current sheet, the magnetic field strength and in-flow speed outside of the bi-pole and the current sheet, the out-flow speed away from the current sheet, and the upward mass flux are written as $B_\mathrm{i}$, $v_\mathrm{i}$, $B_\mathrm{e}$, $v_\mathrm{e}$, $v$, and $F$, respectively. Finally, $\sigma$ denotes the electric conductivity, $m_\mathrm{p}$ is the proton mass, $k$ is Boltzmann's constant, and $c$ is the speed of light.

Assuming iso-thermal reconnection, the VAL-C (\citealt{Vernazza81}) model can be used to give $\sigma=9.9\times10^{10}$ s$^{-1}$, $n_\mathrm{i}=2.1\times10^{15}$ cm$^{-3}$, $\Lambda\approx100$ km, and $T=T_\mathrm{i}=4200$ K (at a height of approximately $500$ km in the photosphere). In addition to this, we allow $v_\mathrm{e}=300$ m$^{-1}$ which was the average in-flow speed measured by \citet{Litvinenko07}. Technically, this quantity could be measured from observations, however, the relatively low-resolution of SDO/HMI magnetograms and the short-lived nature of analysed features would likely cause errors in any estimate for the events of interest to this article. Finally, flux build-up is considered such that:
\begin{equation}
r=\frac{R}{l}=B_\mathrm{i}v_\mathrm{i}=B_\mathrm{e}v_\mathrm{e}.
\end{equation}
As the VAL-C model is unlikely to accurately portray the atmospheric conditions co-spatial to an EB (but is suitable for an initial study as presented here), it would be of interest to conduct such a study in the future alongside numerical simulations of photospheric magnetic restructuring (such as those presented in \citealt{Nelson13b}). 

\begin{table*}
\begin{center}
\begin{tabular}{|c|c|c|c|c|c|c|c|c|}
\hline
&&\multicolumn{2}{l|}{\textbf{Measured Variables}} & \multicolumn{3}{l|}{\textbf{Calculated Variables}}\\
\hline
AR & Event & $R$ (Mx s$^{-1}$) & $l$ (cm) & $B_\mathrm{i}$ (G) & $v$ (km s$^{-1}$) & $H_\mathrm{jet}$ (km) & $E_\mathrm{i}$ (erg) \\
\hline
11579 & MMF 1 & $4.1\times10^{14}$  & $3.1\times10^8$ & $156$ & $5.5$ & $137$ & $1.0\times10^{24}$ \\
& MMF 2 & $6.4\times10^{14}$ & $2.1\times10^{8}$ & $249$ & $6.8$ & $232$ &  $1.2\times10^{24}$ \\
\hline
11765 & Can. Event 1 & $1.6\times10^{15}$  & $2.6\times10^8$ & $362$ & $7.5$ & $406$ & $2.3\times10^{24}$ \\
& Can. Event 2 & $2.4\times10^{15}$ & $2.5\times10^{8}$ & $457$ & $7.8$ & $599$ & $2.9\times10^{24}$ \\
& Can. Event 3 & $1.6\times10^{15}$ & $2.3\times10^{8}$ & $386$ & $7.6$ & $451$ & $2.2\times10^{24}$ \\
\hline
11850 & Event 1 & $1.8\times10^{15}$  & $2.2\times10^8$ & $421$ & $7.7$ & $520$ & $2.3\times10^{24}$ \\
& Event 2 & $1.5\times10^{15}$ & $2.6\times10^{8}$ & $351$ & $7.5$ & $384$ & $2.2\times10^{24}$ \\
& Event 3 & $3.7\times10^{15}$  & $3.5\times10^8$ & $480$ & $7.8$ & $654$ & $4.2\times10^{24}$ \\
& Event 4 & $1.5\times10^{15}$ & $2.2\times10^{8}$ & $382$ & $7.6$ & $443$ & $2.1\times10^{24}$ \\
& Event 5 & $5.0\times10^{14}$  & $1.3\times10^8$ & $282$ & $7.1$ & $276$ & $8.8\times10^{23}$ \\
& Event 6 & $1.9\times10^{15}$ & $2.2\times10^{8}$ & $433$ & $7.7$ & $545$ & $2.4\times10^{24}$ \\
\hline
\end{tabular}
\caption{The measured input values and selected computed output values for each of the cancellation events analysed in this article for the isothermal ($4200$ K) model. All jet heights are calculated using an estimated magnetic field strength of $200$ G.}
\label{Canevetabiso}
\end{center}
\end{table*}

It is then possible to use Eqs.~(6)-(13) from \citet{Litvinenko07} to estimate the further variables required to complete this system. It is interesting to consider that numerous predictions of EBs have suggested that a temperature enhancement occurs at the potential reconnection site (see, for example, \citealt{Kitai83}, \citealt{Fang06} in addition to the recent work by \citealt{Vissers15} and \citealt{Rutten16}). It would, therefore, be of interest to modify this approximation slightly such that iso-thermal reconnection is not demanded. We model this basically by including a further condition that:
\begin{equation}
T=T_\mathrm{i}+\Gamma,
\end{equation}
where $\Gamma$ is a non-zero, positive constant. It should be noted that this deviation from the iso-thermal model does create issues with the Equation of State, however, in the interest of this work, we shall limit $\Gamma$ to be small such that any errors are minimalised. Eq.~(6) and Eq.~(11) from \citet{Litvinenko07} can then be rewritten as:
\begin{equation}
v_\mathrm{i}^3=\frac{c^2r}{4\pi\Lambda\sigma\left(4\pi{m}_\mathrm{p}n_\mathrm{i}\right)^\frac{1}{2}}\left(\frac{T_\mathrm{i}}{T}+\frac{r^2}{8\pi{k}n_\mathrm{i}Tv_\mathrm{i}^2}\right)^\frac{1}{2}
\label{nonisoinflow}
\end{equation}
and
\begin{equation}
n=\frac{n_\mathrm{i}T_\mathrm{i}}{T}+\frac{r^2}{8\pi{k}Tv_\mathrm{i}^2}.
\label{dennew}
\end{equation}

We are now in a position to define three quantities of interest to this study, namely the up-flow speed of the ejected plasma, the height of the corresponding jet, and the energy contained within the magnetic field. The up-flow speed is simply calculated using Eq.~(12) from \citet{Litvinenko07} which is written as:
\begin{equation}
v=\frac{B_\mathrm{i}}{(4\pi{m}_\mathrm{p}n)^{1/2}}
\label{upflow}
\end{equation}
where each of the above variables is returned by the model for a given cancellation rate and current sheet length (cross-sectional diameter).

Secondly, \citet{Takasao13} estimated the maximum height of a jet from a reconnecting region assuming all of the kinetic energy were converted into potential energy. This was written as:
\begin{equation}
H_\mathrm{jet}\sim\frac{\Lambda}{\beta}
\label{jetextent}
\end{equation}
where $\beta$ is the plasma beta defined as:
\begin{equation}
\beta=\frac{B_\mathrm{est}^2}{(8\pi{n}kT)}.
\label{beta}
\end{equation}
Using the output from the \citet{Litvinenko07} model, it is possible to estimate this quantity for each cancellation feature. We shall use an estimate of $B_\mathrm{est}$=$200$ G in order to compare between events and with observational results from \citet{Watanabe11} and \citet{Nelson15}.

Finally, it is possible to estimate the magnetic energy contained within the current sheet from the \citet{Litvinenko07} model using the relation:
\begin{equation}
E_\mathrm{i}=\frac{B_\mathrm{i}^2}{8\pi}\times{V}
\end{equation}
where $V=l\times2a\times2\Lambda$. This value should be of interest to compare to predicted EB (see, for example, \citealt{Georgoulis02}) and IRIS burst (\citealt{Peter14}) energies from within the literature.

\subsection{AR $11579$}

Considering MMF $1$, the measured and estimated parameters are $R=4.1\times10^{14}$ Mx s$^{-1}$ and $l=3.1\times10^8$ cm. When included in the isothermal ($T=4200$ K) Eqs.(6)-(13) from \citet{Litvinenko07}, output values of $v_\mathrm{i}\approx85$ m s$^{-1}$, $B_\mathrm{e}\approx44$ G,  $B_\mathrm{i}\approx156$ G, $n\approx3.8\times10^{15}$ cm$^{-3}$, $v\approx5.5$ km s$^{-1}$, and $F\approx6.6\times10^{14}$ g hr$^{-1}$ are obtained. The input values of $R=6.4\times10^{14}$ Mx s$^{-1}$ and $l=2.1\times10^{8}$ cm for MMF $2$ return values of $v_\mathrm{i}\approx122$ m s$^{-1}$, $B_\mathrm{e}\approx101$ G,  $B_\mathrm{i}\approx249$ G, $n\approx6.4\times10^{15}$ cm$^{-3}$, $v\approx6.8$ km s$^{-1}$, and $F\approx6.5\times10^{14}$ g hr$^{-1}$. Each of the estimated values for both events are comparable to those calculated by \citet{Litvinenko07}. Interestingly, the up-flow velocities of $5.6$ km s$^{-1}$ and $6.8$ km s$^{-1}$ are comparable to the lower limit of velocities measured close to the limb by \citet{Watanabe11} and \citet{Nelson15}. The maximum jet extentions of $137$ km and $232$ km (estimated from Eq.~\ref{jetextent} with an estimated magnetic field of $200$ G) are, however, much lower than measured values for EBs. In addition to this, the potential magnetic energies available for conversion at the current sheet is estimated to be of the correct order to account for the radiative losses of weak EBs at $E_\mathrm{i}\approx10^{24}$ ergs (as was discussed by \citealt{Georgoulis02}). Assuming an efficient conversion of this quantity to radiative and kinetic energy over the course of one second, a maximum energy release over the lifetime of a typical EB (approximatetly $10$ minutes), could be as high as $E_\mathrm{max}\approx6\times10^{26}$ ergs. A summary of both events is included in the appropriate row of Table~\ref{Canevetabiso}.  

\begin{figure}
{\includegraphics[scale=0.33,trim={0 1cm 0 2cm}]{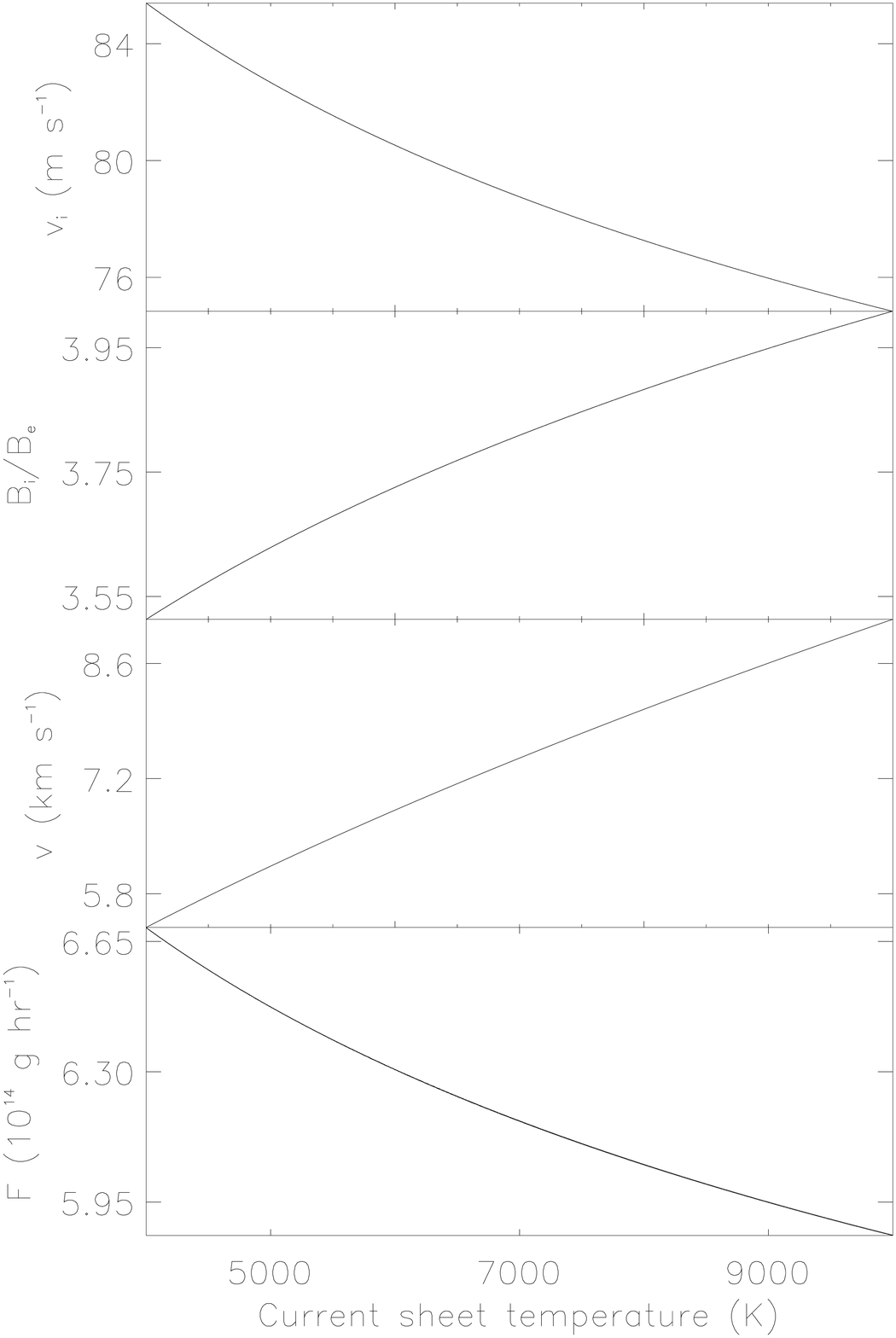}}
\caption{A representative example of the dependence of the in-flow velocity, ratio of current sheet magnetic field to external magnetic field, out-flow velocity and mass flux rate on current sheet temperature for MMF $1$.}
\label{temp_anal}
\end{figure}

In Fig.~\ref{temp_anal}, a number of parameters are plotted with respect to temperature at the current sheet for MMF $1$. Both the in-flow velocity and mass flux rate decrease with increased temperature at the current sheet, whereas the ratio of the current sheet magnetic field to the incoming magnetic field and the up-flow velocity both increase at higher temperatures. Interestingly, only a large increase in temperature at the site of the current sheet (around $10000$ K, which is higher than previous semi-empirical estimates of heating within EBs but of the same order of magnitude as the temperatures estimated by \citealt{Vissers15}) raises the predicted out-flow velocity for this cancellation feature to the average vertical propagation speed of EBs observed at the limb by \citealt{Nelson15}. The upward mass flux predicted for this event is an order of magnitude smaller than that predicted by both \citet{Litvinenko99b} and \citet{Litvinenko07} and the time required to fill a filament with typical mass around $5\times10^{16}$ g (suggested by \citealt{Litvinenko07}) is on the order of $100$ hours (two orders of magnitude larger than the length of cancellation). Of course, small-scale cancellation features with lifetimes of less than one hour are unlikely to prove essential for the formation of filaments; however, using the estimates of \citet{Bong14}, the mass fluxes returned by this model would account for the required filling of surges within one minute under ideal conditions ({\it i.e.}, no gravity etc.).

\subsection{ARs $11765$ \& $11850$}

The values measured for Can. Event $1$ within AR $11765$ of $R=1.6\times10^{15}$ Mx s$^{-1}$ and $l=2.6\times10^8$ cm, return slightly larger output values (than those calculated for the MMFs within AR $11579$) of $v_\mathrm{i}\approx164$ m s$^{-1}$, $B_\mathrm{e}\approx192$ G, $B_\mathrm{i}\approx 351$ G, $n\approx10^{15}$ cm$^{-3}$, $v\approx7.5$ km s$^{-1}$, $F\approx1.1\times10^{15}$ g hr$^{-1}$, and $H_\mathrm{jet}\approx406$ km (again for an estimated field strength of $200$ G). Input values of $R=2.4\times10^{15}$ Mx s$^{-1}$ and $l=2.5\times10^8$ cm corresponding to Can. Event $2$ return outputs of $v_\mathrm{i}\approx210$ m s$^{-1}$, $B_\mathrm{e}\approx320$ G, $B_\mathrm{i}\approx 457$ G, $n\approx1.6\times10^{16}$ cm$^{-3}$, $v\approx7.8$ km s$^{-1}$, $F\approx1.3\times10^{15}$ g hr$^{-1}$, and $H_\mathrm{jet}\approx599$ km. Both the up-flow velocities and jet heights are closer to those observed within EBs by both \citet{Watanabe11} and \citet{Nelson15}. Can. Event $3$ returns similar values of $v_\mathrm{i}\approx179$ m s$^{-1}$, $B_\mathrm{e}\approx231$ G, $B_\mathrm{i}\approx 387$ G, $n\approx1.2^{16}$ cm$^{-3}$, $v\approx7.6$ km s$^{-1}$, $F\approx1.0\times10^{15}$ g hr$^{-1}$, and $H_\mathrm{jet}\approx451$ km. Interestingly, the potential energies contained within these events are an order of $2$ larger than those predicted for MMFs $1$ and $2$ from AR $11579$ at around $2.5\times10^{24}$ ergs. If this quantity is converted away from magnetic energy over the course of one second and then re-supplied by the surrounding bi-pole, these cancellation rates would be capable of contributing enough radiative energy to account for EBs with $E_\mathrm{max}\approx1.2\times10^{27}$ ergs (according to the estimates suggested by \citealt{Georgoulis02}). The output values returned by the model for the six cancellation features observed within AR $11850$ are similar to those returned in AR $11579$. A brief description of each of these bi-poles is included in Table.~\ref{Canevetabiso}.

\section{Discussion}

In this article, eleven well-defined cancellation events have been analysed. The overall aim of this research was to further understand the potential role of cancellation in UV burst formation. Initially, two small bi-polar magnetic flux elements surrounding a sunspot in AR $11579$ were identified and tracked within SDO/HMI data using YAFTA (see \citealt{Welsch03}). Co-aligned imaging data (collected by the SDO/AIA and DST/IBIS instruments) sampling the $1600$ \AA\ and $1700$ \AA\ filters and H$\alpha$ line wings were then visually inspected to confirm or deny the formation of any burst feature. The total magnetic field strength of the representative polarities were also determined by YAFTA and plotted through time (see the bottom frame of Fig.~\ref{caneve_anal}) in order to interpret whether any cancellation was apparent. Both of the events analysed in AR $11579$ exhibited significant cancellation through their lifetimes at rates on the order of $10^{14}$ Mx s$^{-1}$.

The H$\alpha$ intensity increases co-spatial to the MMFs had several typical properties of EBs, including emissions in excess of $150$ \% of the background intensity in the wings of the H$\alpha$ line profile, areas close to $1$\arcsec$\times1$\arcsec, and lifetimes on the order of minutes (see, for example, \citealt{Georgoulis02}, \citealt{Watanabe11}). The SDO/AIA UV channels, though, displayed no evidence of localised brightening in conjuction with the H$\alpha$ wing intensities potentially hinting that these events are magnetic concentrations and not EBs. Interestingly, the peaks of intensity for the H$\alpha$ features co-spatial to MMF $1$ and MMF $2$ were co-temporal to the greatest measured cancellation. Fig.~\ref{caneve_anal} plots representative frames of the magnetic and spectral data during the H$\alpha$ event (top three frames) and the evolution of the magnetic field strength through time (bottom frame). Determining whether the increased cancellation rate at that time contributed to the formation of the increased intensity within this localised region or not is beyond the capabilities of this dataset and would require a larger statistical sample of MMFs inferred from magnetic data collected by a higher-resolution instrument (when compared to SDO/HMI) such as the SST/CRISP. Future work should aim to complete such a study. It is possible that the H$\alpha$ intensity increases co-spatial to the MMFs presented here are similar to the events analysed by \citet{Nindos98}; however, our results are inconclusive as to whether these features are indeed EBs or the more benign magnetic concentrations.

In addition to these MMFs, three cancellation features within AR $11765$ (previously researched by \citealt{Hong14}) and six from AR $11850$ (including the four IRIS hot explosions discussed by \citealt{Peter14}) were analysed. Seven of these bi-poles were, at some point during their lifetimes, co-spatial to UV burst-like behaviour. In Fig~\ref{evolution_anal}, the evolution of one bi-pole from within AR $11765$ was plotted through time depicting the spatial relationship between the cancellation and the UV burst. In addition to this, eight of the bi-poles displayed measureable cancellation rates of the order $10^{15}$ Mx s$^{-1}$ (four of which are plotted for reference in the top rows of Fig.~\ref{can_11765} and Fig.~\ref{can_11850}), with the ninth event cancelling at a rate of $5\times10^{14}$ Mx s$^{-1}$. For the seven events co-spatial to bursts, it was found that the periods of increased UV intensity in the SDO/AIA $1600$ \AA\ and $1700$ \AA\ channels were co-temporal to sustained cancellation (as plotted in Fig.~\ref{can_11765} and Fig.~\ref{can_11850}). The morphology of these bursts was similar to UV intensity enhancements measured by \citet{Vissers15} co-spatial to both EBs and FAFs potentially hinting at the formation of the phenomena studied here. Which category these events could possibly fall into is currently unknown (without co-temporal H$\alpha$ data). No discernible properties ({\it e.g.}, cancellation rates) were found to differentiate or explain the two non-burst related cancellation features within these datasets, however, it is known that an IRIS hot explosion occurred co-spatial to one of these bi-poles (\citealt{Peter14}). It is possible that this event corresponds to a subset of IRIS bursts with no photospheric component.

The measured properties of all eleven cancellation features were then applied to the magnetic reconnection model of \citet{Litvinenko99a} and \citet{Litvinenko07}. Specifically of interest were the upward flow velocity, the maximum jet height, and the energy estimates at the site of the hypothesised reconnection which can be directly compared to the measurements obtained at the solar limb (by, for example, \citealt{Nelson15}). Assuming an iso-thermal reconnection site, all features analysed here returned estimated up-flow velocities of between $5$ km s$^{-1}$ and $8$ km s$^{-1}$, which are within one standard deviation of the mean upward flow velocity found in \citet{Nelson15}. The maximum jet heights (assuming a magnetic field of around $200$ G) for events with cancellation rates over $10^{15}$ Mx s$^{-1}$ were also similar to the measured heights within the literature on EBs. The maximum energies contained within the magnetic field are comparable to the radiative energies measured within EBs at approximately $1.2\times10^{27}$ ergs (as has been discussed by \citealt{Georgoulis02}), although they are lower than those suggested for IRIS hot explosions (\citealt{Peter14}).

If iso-thermal reconnection is not assumed, the up-flow velocity increases. Specifically, enhancements in temperature corresponding to those estimated by semi-empirical modelling of EBs (see, for example, \citealt{Kitai83}, \citealt{Fang06}) appear to improve the fit of this model to observed measurements of upward flow speed (such as \citealt{Watanabe11}, \citealt{Nelson15}); however, this model does not explain the larger heating within EBs and IRIS hot explosions recently reported by both \citet{Peter14} and \citet{Vissers15}. How such heating would occur is still unknown. It would be of interest to conduct a similar study as presented here with a wider variety of data including both EBs observed co-spatial to spectra sampled by the IRIS instrument and also co-spatial to surges. Such research should allow for an accurate estimation of the cancellation rate co-spatial to both of these features, potentially allowing further inferences to be made about the temperature structures of these events.

The apparently sporadic relationship between UV bursts, surges, and cancellation features is both confusing and intriguing. Within the literature, numerous authors have discussed the potential relationship between EBs and surges (see, {\it e.g.}, \citealt{Rust68}, \citealt{Roy73}, \citealt{Madjarska09}); however, various datasets have also been analysed showing no links between these features ({\it e.g.}, \citealt{Vissers13}, \citealt{Nelson15}). Whether specific conditions must be achieved within a multi-stage process for an UV burst to drive a surge or whether these events are unrelated other than by an occasional co-spatial occurrence is still to be understood. \citet{Platov73} suggested that an increase in the strength of a bi-polar region (up to a factor of $25$) would be required within $5$ minutes in order for a surge to form. The eleven bi-poles analysed here do not show such changes, nor a link to surge events. The cancellation rates  (\citealt{Roy73}) and mass fluxes (\citealt{Bong14}) measured here, though, are similar to those potentially required for surge formation. It is possible that only a subset of UV bursts which correspond to higher cancellation rates are linked to surges. This relationship should be tested in future work.

\section*{acknowledgements}
The authors thank the anonymous reviewer for their comments which improved the document. Research at the Armagh Observatory is grant-aided by the N. Ireland Dept. of Communities. The authors are thankful to the UK Science and Technology Facilities Council (STFC) UK for support received via a studentship and T\&S via PATT. Data in this publication were obtained with the facilities of and with help from the staff of the National Solar Observatory, which is operated by the Association of Universities for Research in Astronomy, Inc. (AURA), under cooperative agreement with the National Science Foundation. Friedrich W\"oger and Kevin Reardon must also be thanked for their assistance with IBIS data reconstruction. SDO/AIA and SDO/HMI data is courtesy of NASA/SDO and the AIA and HMI science teams. R.E. is also grateful to NSF, Hungary (OTKA, Ref. No. K83133). CJN thanks ISSI Bern for the support to the team ”Solar UV bursts – a new insight to magnetic reconnection”.

\bibliographystyle{mn2e}
\bibliography{EBAnalytics.bib}

\begin{thebibliography}{}

\bibitem[\protect\citeauthoryear{{Bong}, {Cho} \& {Yurchyshyn}}{{Bong}
  et~al.}{2014}]{Bong14}
{Bong} S.-C.,  {Cho} K.-S.,    {Yurchyshyn} V.,  2014, Journal of Korean
  Astronomical Society, 47, 311

\bibitem[\protect\citeauthoryear{{Cavallini} \& {IBIS Team}}{{Cavallini} \&
  {IBIS Team}}{2004}]{Cavallini04}
{Cavallini} F.,  {IBIS Team} 2004, in American Astronomical Society Meeting
  Abstracts \#204 Vol.~36 of Bulletin of the American Astronomical Society,
  {IBIS: Instrument Description and First Results}.
p.~710

\bibitem[\protect\citeauthoryear{{Chae}, {Moon} \& {Park}}{{Chae}
  et~al.}{2003}]{Chae03}
{Chae} J.,  {Moon} Y.-J.,    {Park} S.-Y.,  2003, Journal of Korean
  Astronomical Society, 36, 13

\bibitem[\protect\citeauthoryear{{de Pontieu}, {McIntosh}, {Hansteen},
  {Carlsson}, {Schrijver}, {Tarbell}, {Title}, {Shine}, {Suematsu}, {Tsuneta},
  {Katsukawa}, {Ichimoto}, {Shimizu} \& {Nagata}}{{de Pontieu}
  et~al.}{2007}]{dePontieu07}
{de Pontieu} B.,  {McIntosh} S.,  {Hansteen} V.~H.,  {Carlsson} M.,
  {Schrijver} C.~J.,  {Tarbell} T.~D.,  {Title} A.~M.,  {Shine} R.~A.,
  {Suematsu} Y.,  {Tsuneta} S.,  {Katsukawa} Y.,  {Ichimoto} K.,  {Shimizu} T.,
     {Nagata} S.,  2007, \pasj, 59, 655

\bibitem[\protect\citeauthoryear{{Ellerman}}{{Ellerman}}{1917}]{Ellerman17}
{Ellerman} F.,  1917, \apj, 46, 298

\bibitem[\protect\citeauthoryear{{Fang}, {Tang}, {Xu}, {Ding} \& {Chen}}{{Fang}
  et~al.}{2006}]{Fang06}
{Fang} C.,  {Tang} Y.~H.,  {Xu} Z.,  {Ding} M.~D.,    {Chen} P.~F.,  2006,
  \apj, 643, 1325

\bibitem[\protect\citeauthoryear{{Georgoulis}, {Rust}, {Bernasconi} \&
  {Schmieder}}{{Georgoulis} et~al.}{2002}]{Georgoulis02}
{Georgoulis} M.~K.,  {Rust} D.~M.,  {Bernasconi} P.~N.,    {Schmieder} B.,
  2002, \apj, 575, 506

\bibitem[\protect\citeauthoryear{{Harvey} \& {Harvey}}{{Harvey} \&
  {Harvey}}{1973}]{Harvey73}
{Harvey} K.,  {Harvey} J.,  1973, \solphys, 28, 61

\bibitem[\protect\citeauthoryear{{Hong}, {Ding}, {Li}, {Fang} \& {Cao}}{{Hong}
  et~al.}{2014}]{Hong14}
{Hong} J.,  {Ding} M.~D.,  {Li} Y.,  {Fang} C.,    {Cao} W.,  2014, \apj, 792,
  13

\bibitem[\protect\citeauthoryear{{Kitai}}{{Kitai}}{1983}]{Kitai83}
{Kitai} R.,  1983, \solphys, 87, 135

\bibitem[\protect\citeauthoryear{{Lemen}}{{Lemen}}{2012}]{Lemen12}
{Lemen} J.~R. e.~a.,  2012, \solphys, 275, 17

\bibitem[\protect\citeauthoryear{{Li} \& {Zhang}}{{Li} \& {Zhang}}{2013}]{Li13}
{Li} X.,  {Zhang} H.,  2013, \apj, 771, 22

\bibitem[\protect\citeauthoryear{{Lim}, {Yurchyshyn} \& {Goode}}{{Lim}
  et~al.}{2012}]{Lim12}
{Lim} E.-K.,  {Yurchyshyn} V.,    {Goode} P.,  2012, \apj, 753, 89

\bibitem[\protect\citeauthoryear{{Litvinenko}}{{Litvinenko}}{1999}]{Litvinenko99a}
{Litvinenko} Y.~E.,  1999, \apj, 515, 435

\bibitem[\protect\citeauthoryear{{Litvinenko}, {Chae} \& {Park}}{{Litvinenko}
  et~al.}{2007}]{Litvinenko07}
{Litvinenko} Y.~E.,  {Chae} J.,    {Park} S.-Y.,  2007, \apj, 662, 1302

\bibitem[\protect\citeauthoryear{{Litvinenko} \& {Martin}}{{Litvinenko} \&
  {Martin}}{1999}]{Litvinenko99b}
{Litvinenko} Y.~E.,  {Martin} S.~F.,  1999, \solphys, 190, 45

\bibitem[\protect\citeauthoryear{{Madjarska}, {Doyle} \& {de
  Pontieu}}{{Madjarska} et~al.}{2009}]{Madjarska09}
{Madjarska} M.~S.,  {Doyle} J.~G.,    {de Pontieu} B.,  2009, \apj, 701, 253

\bibitem[\protect\citeauthoryear{{Nelson}, {Scullion}, {Doyle}, {Freij} \&
  {Erd{\'e}lyi}}{{Nelson} et~al.}{2015}]{Nelson15}
{Nelson} C.~J.,  {Scullion} E.~M.,  {Doyle} J.~G.,  {Freij} N.,
  {Erd{\'e}lyi} R.,  2015, \apj, 798, 19

\bibitem[\protect\citeauthoryear{{Nelson}, {Shelyag}, {Mathioudakis}, {Doyle},
  {Madjarska}, {Uitenbroek} \& {Erd{\'e}lyi}}{{Nelson}
  et~al.}{2013}]{Nelson13b}
{Nelson} C.~J.,  {Shelyag} S.,  {Mathioudakis} M.,  {Doyle} J.~G.,  {Madjarska}
  M.~S.,  {Uitenbroek} H.,    {Erd{\'e}lyi} R.,  2013, \apj, 779, 125

\bibitem[\protect\citeauthoryear{{Nindos} \& {Zirin}}{{Nindos} \&
  {Zirin}}{1998}]{Nindos98}
{Nindos} A.,  {Zirin} H.,  1998, \solphys, 182, 381

\bibitem[\protect\citeauthoryear{{Parker}}{{Parker}}{1957}]{Parker57}
{Parker} E.~N.,  1957, \jgr, 62, 509

\bibitem[\protect\citeauthoryear{{Peter}, {Tian}, {Curdt}, {Schmit}, {Innes} \&
  {De Pontieu}}{{Peter} et~al.}{2014}]{Peter14}
{Peter} H.,  {Tian} H.,  {Curdt} W.,  {Schmit} D.,  {Innes} D.,    {De Pontieu}
  B.~e.,  2014, Science, 346, 1255726

\bibitem[\protect\citeauthoryear{{Platov}, {Somov} \& {Syrovatskii}}{{Platov}
  et~al.}{1973}]{Platov73}
{Platov} Y.~V.,  {Somov} B.~V.,    {Syrovatskii} S.~I.,  1973, \solphys, 30,
  139

\bibitem[\protect\citeauthoryear{{Qiu}, {Ding}, {Wang}, {Denker} \&
  {Goode}}{{Qiu} et~al.}{2000}]{Qiu00}
{Qiu} J.,  {Ding} M.~D.,  {Wang} H.,  {Denker} C.,    {Goode} P.~R.,  2000,
  \apjl, 544, L157

\bibitem[\protect\citeauthoryear{{Reid}, {Mathioudakis}, {Doyle}, {Scullion},
  {Nelson}, {Henriques} \& {Ray}}{{Reid} et~al.}{2016}]{Reid16}
{Reid} A.,  {Mathioudakis} M.,  {Doyle} J.~G.,  {Scullion} E.,  {Nelson} C.~J.,
   {Henriques} V.,    {Ray} T.,  2016, \apj, 823, 110

\bibitem[\protect\citeauthoryear{{Reid}, {Mathioudakis}, {Scullion}, {Doyle},
  {Shelyag} \& {Gallagher}}{{Reid} et~al.}{2015}]{Reid15}
{Reid} A.,  {Mathioudakis} M.,  {Scullion} E.,  {Doyle} J.~G.,  {Shelyag} S.,
   {Gallagher} P.,  2015, \apj, 805, 64

\bibitem[\protect\citeauthoryear{{Rouppe van der Voort}, {Rutten} \&
  {Vissers}}{{Rouppe van der Voort} et~al.}{2016}]{vanderVoort16}
{Rouppe van der Voort} L.~H.~M.,  {Rutten} R.~J.,    {Vissers} G.~J.~M.,  2016,
  ArXiv e-prints

\bibitem[\protect\citeauthoryear{{Roy}}{{Roy}}{1973}]{Roy73}
{Roy} J.-R.,  1973, \solphys, 32, 139

\bibitem[\protect\citeauthoryear{{Rust}}{{Rust}}{1968}]{Rust68}
{Rust} D.~M.,  1968, in {Kiepenheuer} K.~O.,  ed., Structure and Development of
  Solar Active Regions Vol.~35 of IAU Symposium, {Chromospheric Explosions and
  Satellite Sunspots}.
p.~77

\bibitem[\protect\citeauthoryear{{Rutten}}{{Rutten}}{2016}]{Rutten16}
{Rutten} R.~J.,  2016, \aap, 590, A124

\bibitem[\protect\citeauthoryear{{Rutten}, {Vissers}, {Rouppe van der Voort},
  {S{\"u}tterlin} \& {Vitas}}{{Rutten} et~al.}{2013}]{Rutten13}
{Rutten} R.~J.,  {Vissers} G.~J.~M.,  {Rouppe van der Voort} L.~H.~M.,
  {S{\"u}tterlin} P.,    {Vitas} N.,  2013, Journal of Physics Conference
  Series, 440, 012007

\bibitem[\protect\citeauthoryear{{Scharmer}}{{Scharmer}}{2006}]{Scharmer06}
{Scharmer} G.~B.,  2006, \aap, 447, 1111

\bibitem[\protect\citeauthoryear{{Scharmer}, {Narayan}, {Hillberg}, {de la Cruz
  Rodr{\'{\i}}guez}, {L{\"o}fdahl}, {Kiselman}, {S{\"u}tterlin}, {van Noort} \&
  {Lagg}}{{Scharmer} et~al.}{2008}]{Scharmer08}
{Scharmer} G.~B.,  {Narayan} G.,  {Hillberg} T.,  {de la Cruz Rodr{\'{\i}}guez}
  J.,  {L{\"o}fdahl} M.~G.,  {Kiselman} D.,  {S{\"u}tterlin} P.,  {van Noort}
  M.,    {Lagg} A.,  2008, \apjl, 689, L69

\bibitem[\protect\citeauthoryear{{Scherrer}, {Schou}, {Bush}, {Kosovichev},
  {Bogart}, {Hoeksema}, {Liu}, {Duvall}, {Zhao}, {Title}, {Schrijver},
  {Tarbell} \& {Tomczyk}}{{Scherrer} et~al.}{2012}]{Scherrer12}
{Scherrer} P.~H.,  {Schou} J.,  {Bush} R.~I.,  {Kosovichev} A.~G.,  {Bogart}
  R.~S.,  {Hoeksema} J.~T.,  {Liu} Y.,  {Duvall} T.~L.,  {Zhao} J.,  {Title}
  A.~M.,  {Schrijver} C.~J.,  {Tarbell} T.~D.,    {Tomczyk} S.,  2012,
  \solphys, 275, 207

\bibitem[\protect\citeauthoryear{{Sheeley} Jr.}{{Sheeley}}{1969}]{Sheeley69}
{Sheeley} Jr. N.~R.,  1969, \solphys, 9, 347

\bibitem[\protect\citeauthoryear{{Shibata}, {Nishikawa}, {Kitai} \&
  {Suematsu}}{{Shibata} et~al.}{1982}]{Shibata82}
{Shibata} K.,  {Nishikawa} T.,  {Kitai} R.,    {Suematsu} Y.,  1982, \solphys,
  77, 121

\bibitem[\protect\citeauthoryear{{Suematsu}, {Shibata}, {Neshikawa} \&
  {Kitai}}{{Suematsu} et~al.}{1982}]{Suematsu82}
{Suematsu} Y.,  {Shibata} K.,  {Neshikawa} T.,    {Kitai} R.,  1982, \solphys,
  75, 99

\bibitem[\protect\citeauthoryear{{Sweet}}{{Sweet}}{1958}]{Sweet58}
{Sweet} P.~A.,  1958, in {Lehnert} B.,  ed., Electromagnetic Phenomena in
  Cosmical Physics Vol.~6 of IAU Symposium, {The Neutral Point Theory of Solar
  Flares}.
p.~123

\bibitem[\protect\citeauthoryear{{Takasao}, {Isobe} \& {Shibata}}{{Takasao}
  et~al.}{2013}]{Takasao13}
{Takasao} S.,  {Isobe} H.,    {Shibata} K.,  2013, \pasj, 65, 62

\bibitem[\protect\citeauthoryear{{Vernazza}, {Avrett} \& {Loeser}}{{Vernazza}
  et~al.}{1981}]{Vernazza81}
{Vernazza} J.~E.,  {Avrett} E.~H.,    {Loeser} R.,  1981, \apjs, 45, 635

\bibitem[\protect\citeauthoryear{{Vissers}, {Rouppe van der Voort} \&
  {Rutten}}{{Vissers} et~al.}{2013}]{Vissers13}
{Vissers} G.~J.~M.,  {Rouppe van der Voort} L.~H.~M.,    {Rutten} R.~J.,  2013,
  \apj, 774, 32

\bibitem[\protect\citeauthoryear{{Vissers}, {Rouppe van der Voort}, {Rutten},
  {Carlsson} \& {De Pontieu}}{{Vissers} et~al.}{2015}]{Vissers15}
{Vissers} G.~J.~M.,  {Rouppe van der Voort} L.~H.~M.,  {Rutten} R.~J.,
  {Carlsson} M.,    {De Pontieu} B.,  2015, \apj, 812, 11

\bibitem[\protect\citeauthoryear{{Watanabe}, {Vissers}, {Kitai}, {Rouppe van
  der Voort} \& {Rutten}}{{Watanabe} et~al.}{2011}]{Watanabe11}
{Watanabe} H.,  {Vissers} G.,  {Kitai} R.,  {Rouppe van der Voort} L.,
  {Rutten} R.~J.,  2011, \apj, 736, 71

\bibitem[\protect\citeauthoryear{{Welsch} \& {Longcope}}{{Welsch} \&
  {Longcope}}{2003}]{Welsch03}
{Welsch} B.~T.,  {Longcope} D.~W.,  2003, \apj, 588, 620

\bibitem[\protect\citeauthoryear{{W{\"o}ger}, {von der L{\"u}he} \&
  {Reardon}}{{W{\"o}ger} et~al.}{2008}]{Woger08}
{W{\"o}ger} F.,  {von der L{\"u}he} O.,    {Reardon} K.,  2008, \aap, 488, 375

\bibitem[\protect\citeauthoryear{{Yang}, {Chae}, {Lim}, {Park}, {Cho},
  {Maurya}, {Song}, {Kim} \& {Goode}}{{Yang} et~al.}{2013}]{Yang13}
{Yang} H.,  {Chae} J.,  {Lim} E.-K.,  {Park} H.,  {Cho} K.,  {Maurya} R.~A.,
  {Song} D.,  {Kim} Y.-H.,    {Goode} P.~R.,  2013, \solphys, 288, 39

\bibitem[\protect\citeauthoryear{{Zachariadis}, {Alissandrakis} \&
  {Banos}}{{Zachariadis} et~al.}{1987}]{Zachariadis87}
{Zachariadis} T.~G.,  {Alissandrakis} C.~E.,    {Banos} G.,  1987, \solphys,
  108, 227

\end{thebibliography}

\bsp

\label{lastpage}

\end{document}